\documentclass[conference]{IEEEtranSPMB}

\IEEEoverridecommandlockouts
\ifCLASSINFOpdf
\else
\fi
\usepackage{amsmath,graphicx,soul,subcaption,balance}
\usepackage[numbers,sort&compress]{natbib}

\usepackage{parskip}
\usepackage{float} 
\usepackage[all]{nowidow}
\usepackage{mathptmx}

\usepackage{multirow} % for enabling multirow and multicol tables

\usepackage{color}
\usepackage{soul,mathrsfs}
\usepackage[utf8]{inputenc}
\usepackage[T1]{fontenc}
\usepackage{array}
\usepackage[hyphens, spaces, obeyspaces]{url}

\usepackage[papersize={8.5in,11in}, left=1in, right=1in, top=1in, bottom=1in]{geometry}
\newcolumntype{L}[1]{>{\raggedright\let\newline\\\arraybackslash\hspace{0pt}}m{#1}}
\newcolumntype{C}[1]{>{\centering\let\newline\\\arraybackslash\hspace{0pt}}m{#1}}
\newcolumntype{R}[1]{>{\raggedleft\let\newline\\\arraybackslash\hspace{0pt}}m{#1}}

\usepackage[singlelinecheck=false,justification=centering]{caption}
\DeclareCaptionLabelFormat{mylabel}{#1 #2.\hspace{0.7ex}}
\renewcommand{\thetable}{\arabic{table}}

\usepackage{colortbl}
\usepackage[dvipsnames]{xcolor}
\definecolor{light-gray}{gray}{0.83}

\newcommand{\spmbtitlefont}{\fontsize{11.0pt}{11.00pt}\selectfont\bf\vspace{0.7em}}
\newcommand{\spmbauthorfont}{\fontsize{11.0pt}{11.0pt}\selectfont\vspace{0em}}
\newcommand{\subparagraph}{}
\usepackage[compact]{titlesec}
\titleformat{\section}
   {\center\normalfont\sc}{\thesection.}{0.7ex}{}
\titlespacing{\section}{0pt}{2ex}{1.5ex}
\titlespacing{\subsection}{0pt}{1.5ex}{1.2ex}
\titlespacing{\subsubsection}{0pt}{1ex}{0.9ex}
\makeatletter
\renewcommand*{\@seccntformat}[1]{\csname the#1\endcsname .\hspace{0.7em}}
\makeatother

\usepackage{fancyhdr,lastpage}
\fancypagestyle{firststyle}
{
   \fancyhf{}
   \fancyfoot[L]{\scriptsize 979-8-3315-7370-6/25/\$31.00 \copyright2025 IEEE}
   \fancyfoot[R]{\scriptsize December 6, 2025}
   \fancyfoot[C]{\scriptsize IEEE SPMB 2025}
}
\usepackage{color}

\usepackage{tikz}
\usetikzlibrary{positioning}
\usepackage{tikzscale}
\usepackage{import}
\usetikzlibrary{quotes,arrows.meta}
\usetikzlibrary{positioning}

\usepackage{Box}
\usepackage{Box_inside_stacked}
\usetikzlibrary{backgrounds,calc,shadings,shapes.arrows,arrows,shapes.symbols,shadows,positioning,decorations.markings,backgrounds,arrows.meta}
\definecolor{darkblue}{HTML}{1f4e79}
\definecolor{lightblue}{HTML}{00b0f0}
\definecolor{salmon}{HTML}{ff9c6b}
\definecolor{dodgerblue}{rgb}{0.12, 0.56, 1.0}
\definecolor{frenchblue}{rgb}{0.0, 0.45, 0.73}
\definecolor{green(pigment)}{rgb}{0.0, 0.65, 0.31}
\definecolor{macaroniandcheese}{rgb}{1.0, 0.74, 0.53}
\definecolor{pansypurple}{rgb}{0.47, 0.09, 0.29}
\definecolor{darkpansypurple}{rgb}{0.4, 0.0, 0.2}
\definecolor{glaucous}{rgb}{0.38, 0.51, 0.71}
\definecolor{hanblue}{rgb}{0.27, 0.42, 0.81}
\definecolor{arylideyellow}{rgb}{0.91, 0.84, 0.42}
\definecolor{newblue}{rgb}{0.56, 0.67, 0.85}
\definecolor{fireenginered}{rgb}{0.81, 0.09, 0.13}
\definecolor{newgreen}{rgb}{0.67, 0.82, 0.57}

\definecolor{newred}{rgb}{0.9, 0.09, 0.1}
\definecolor{newblue}{rgb}{0.56, 0.67, 0.85}
\definecolor{brown}{rgb}{0.6274509804, 0.3215686275, 0.1764705882}
\definecolor{orange}{rgb}{1, 0.9019607843, 0.3921568627}
\definecolor{purple}{rgb}{0.862745098, 0.4705882353, 0.9019607843}
\usepackage{caption}
\usepackage{stackengine}
\usepackage{hyperref}
\hypersetup{hidelinks}
\usetikzlibrary{calc}

\usepackage{array}
\usepackage{booktabs}
\usepackage{multirow}
\usepackage{siunitx}
\usepackage{arydshln}
\makeatletter
\def\adl@drawiv#1#2#3{%
        \hskip.5\tabcolsep
        \xleaders#3{#2.5\@tempdimb #1{1}#2.5\@tempdimb}%
                #2\z@ plus1fil minus1fil\relax
        \hskip.5\tabcolsep}
\newcommand{\cdashlinelr}[1]{%
  \noalign{\vskip\aboverulesep
           \global\let\@dashdrawstore\adl@draw
           \global\let\adl@draw\adl@drawiv}
  \cdashline{#1}
  \noalign{\global\let\adl@draw\@dashdrawstore
           \vskip\belowrulesep}}
\newcommand{\cdashlinelrdotted}[1]{%
  \noalign{\vskip\aboverulesep
           \global\let\@dashdrawstore\adl@draw
           \global\let\adl@draw\adl@drawiv}
  \cdashline{#1}[.4pt/1pt]
  \noalign{\global\let\adl@draw\@dashdrawstore
           \vskip\belowrulesep}}
\makeatother

\newcolumntype{C}{>{$}c<{$}}

\AtBeginDocument{%
  \heavyrulewidth=.08em
  \lightrulewidth=.05em
  \cmidrulewidth=.03em
  \belowrulesep=.65ex
  \belowbottomsep=0pt
  \aboverulesep=.4ex
  \abovetopsep=0pt
  \cmidrulesep=\doublerulesep
  \cmidrulekern=.5em
  \defaultaddspace=.5em
}

\definecolor{darkblue}{HTML}{1f4e79}
\definecolor{lightblue}{HTML}{00b0f0}
\definecolor{salmon}{HTML}{ff9c6b}
\definecolor{dodgerblue}{rgb}{0.12, 0.56, 1.0}
\definecolor{frenchblue}{rgb}{0.0, 0.45, 0.73}
\definecolor{green(pigment)}{rgb}{0.0, 0.65, 0.31}
\definecolor{macaroniandcheese}{rgb}{1.0, 0.74, 0.53}
\definecolor{arylideyellow}{rgb}{0.91, 0.84, 0.42}
\definecolor{pansypurple}{rgb}{0.47, 0.09, 0.29}
\definecolor{glaucous}{rgb}{0.38, 0.51, 0.71}
\definecolor{hanblue}{rgb}{0.27, 0.42, 0.81}
\definecolor{newblue}{rgb}{0.56, 0.67, 0.85}
\definecolor{newgreen}{rgb}{0.67, 0.82, 0.57}
\definecolor{fireenginered}{rgb}{0.81, 0.09, 0.13}

\title{\spmbtitlefont Retrieving Filter Spectra in CNN for Explainable Sleep Stage Classification
{\vspace{-2.4\baselineskip}
}
}
    \author{\spmbauthorfont\IEEEauthorblockN{
    S.~Goerttler\textsuperscript{\it 1,2}, 
    Y.~Wang\textsuperscript{\it 2}, 
    F.~He\textsuperscript{\it 1} and 
    M.~Wu\textsuperscript{\it 2}
    }
    \vspace{0.9em}
    \IEEEauthorblockA{\spmbauthorfont 
        1. Centre for Computational Science and Mathematical Modelling, Coventry University, Coventry, UK \\
        2. Institute for Infocomm Research, A*STAR, Singapore \\
        goerttlers@uni.coventry.ac.uk
    }
}

\newcommand{\PaperTitleSummary}{S.\ Goerttler, et al.: Retrieving Filter Spectra ...}

\begin{document}

\IEEEaftertitletext{}
\maketitle

%------------------------------------------------------------------------
\begin{abstract}

Despite significant advances in deep learning-based sleep stage classification, the clinical adoption of automatic classification models remains slow. One key challenge is the lack of explainability, as many models function as black boxes with millions of parameters. In response, recent work has increasingly focussed on enhancing model explainability. This study contributes to these efforts by introducing an explainability tool for spectral processing of individual EEG channels. Specifically, this tools retrieves the filter spectrum of low-level convolutional feature extraction and compares it with the classification-relevant spectral information in the data. We apply our tool on the EEGNet and MSA-CNN models using the ISRUC-S3 and Sleep-EDF-20 datasets. The tool reveals that spectral processing plays a significant role in the lower frequency bands. In addition, comparing the correlation between filter spectrum and data-derived spectral information with univariate performance indicates that the model naturally prioritises the most informative channels in a multimodal setting. We specify how these insights can be leveraged to enhance model performance. The code for the filter spectrum retrieval and its analysis is available at \url{https://github.com/sgoerttler/MSA-CNN}.
\end{abstract}


%------------------------------------------------------------------------
%{\textbf{\textit{keywords:}}} Keywords are optional.
%%%%%%%%%%%%%%%%%%%%%%%%%
\IEEEpeerreviewmaketitle    % do not delete
\thispagestyle{firststyle}  % do not delete 
%%%%%%%%%%%%%%%%%%%%%%%%%
%------------------------------------------------------------------------
\section{Introduction}
\label{sec:introduction}
Sleep stage classification is essential for assessing sleep and diagnosing sleep disorders using polysomnography~\cite{rundo2019polysomnography}. With advances in machine learning, deep learning-based sleep stage classification has garnered significant attention~\cite{zhang2024review}. However, despite promising results, its clinical adoption remains limited due to challenges in validation, professional integration, and explainability~\cite{yue2024research}. 

Explainability is particularly crucial for regulatory compliance and building trust among clinicians and patients while advancing model development~\cite{rasheed2022explainable}.
As a consequence, explainability of sleep stage classification models has gained significant traction in recent years.
For instance, gradient-weighted class activation mapping has been used for sample-specific post-hoc explanation of input importance~\cite{vaquerizo2023explainable,dutt2023sleepxai}, as well as for highlighting weighted attention~\cite{liu2023micro}.
Furthermore, self-attention has been utilised to explain input importance~\cite{adey2024exploration} and even temporal interactions~\cite{goerttler2025msacnnlightweightmultiscalecnn} in individual samples.
While these approaches demonstrate sample-specific explanations, model explainability can also be global at the dataset level.
For example, surrogate ablation techniques have been used to assess the contribution of individual modalities globally~\cite{ellis2021explainable}.
Note that despite this abundance of available explainability techniques, each method generally captures only specific aspects of model behaviour and is often dependent on the model architecture.
%Despite this abundance of explainability tec, these explainability techniques each only capture certain aspects of model reasoning and may yield differing insights depending on the data modality and architecture.
% These approaches demonstrate local, sample-specific explanations, but these global explainability methods, such as surrogate ablation techniques, have been used to assess the contribution of individual modalities~\cite{ellis2021explainable}, while ignoring the explanality in multivariate signals.
%An interpretable sleep stage classification model can also be achieved by integrating deep learning models with more interpretable, knowledge-driven models~\cite{shao2024reliable}.

%, while also supporting researchers in refining models

%Here
%Interpretability can also 
%, itself part of the model, has been used to explain 
%can be used  weights can be used 
%To explain modality importance globally, 
%Furthermore, 
%Model complexity is also closely tied to explainability, which is a
%critical factor for the clinical deployment of sleep stage classification models. 
%To address this, r
%Recent studies have employed ; yet, to the best of our knowledge, explanations for transformer-based attention remain absent. 
%Alongside model complexity, explainability is another 

%or integrated deep learning models with interpretable, knowledge-driven models~\cite{shao2024reliable}.

This work aims to extend existing efforts by introducing a method for explaining spectral information processing globally in a multivariate context. Our approach can be applied to models that leverage temporal convolutions to extract spectral and morphological features from the raw signal, which are common in EEG signal classification~\cite{craik2019deep}.
%
%understanding the role of each channel for spectral information processing in the model.
In particular, our method retrieves the frequency spectrum of filters in the first convolutional layer. This layer may comprise several pathways to cover the multiple spectral scales relevant in neurophysiology~\cite{supratak2017deepsleepnet,goerttler2025msacnnlightweightmultiscalecnn}, in which case we employ a unification assignment matrix to integrate these pathways.
Further, we correlate the extracted filter spectrum with the classification-relevant spectral information in the data, which provides insight into the extent of spectral information processing. Lastly, we compare these correlations with the single-channel performance.

We evaluate the proposed method on the EEGNet introduced by Lawhern \textit{et al.}~\cite{lawhern2018eegnet} as well as our previously introduced Multi-Scale and Attention Convolutional Neural Network (MSA-CNN), using two datasets. Both models are multivariate, CNN-based, and designed to capture a broad spectral range from 0\,Hz to around 50\,Hz. 

\begin{figure*}[tbh]
    \resizebox{1.005\textwidth}{!}{%
    \input{graphics/networks.tikz}%
    }
    \caption{Overview of proposed filter spectrum retrieval model for uni- (a--b) and multi-scale (c--e) configurations. (a) Trained uni-scale CNN model, such as the EEGNet~\cite{lawhern2018eegnet}. The first convolutional layer extracts temporal features from the raw signals, followed by higher-order 
    %feature extraction and classification 
    modules. 
    %%Only the trained weights of the first layer are used to retrieve the filter spectrum. 
    (b) Proposed uni-scale filter spectrum retrieval using the trained convolutional filter weights.
    %are Fourier-transformed, and the amplitudes of the Fourier components are averaged across filters to produce the final filter spectrum.
    %e first convolutional layer, comprising four multi-scale convolutions,
    (c) Trained multi-scale CNN model, such as the MSA-CNN~\cite{goerttler2025msacnnlightweightmultiscalecnn}. The first layer comprises temporal convolutions on multiple scales, which are aggregated before being passed to the higher-order modules.
    %. The multi-scale feature maps are aggregated before being passed to the higher-order feature extraction and classification modules.
     (d) The multi-scale filter spectrum retrieval conducts the steps in b separately for each scale before the scales are averaged using the unification assignment matrix.
     %Similar to b, the Fourier component amplitudes of the trained filter weights are averaged across filters, but separately for each scale. Multiplication with $\mathbf{S}_\mathrm{unif}$ yields the final filter spectrum.
     (e) The unification assignment matrix $\mathbf{S}_\mathrm{unif}$ transforms scale-based frequency to unified frequency.
     %and can be computed using the pooling values of the multi-scale convolution.
     }
    \label{fig:method}
\end{figure*}

\section{Methodology}
\label{sec:methods}
\subsection{Method Overview}
The principal idea of our proposed method is to explain channel-wise spectral processing within the first convolutional feature extraction layer. 
This layer can capture both spectral features (e.g., EEG oscillations) and morphological features (e.g., waveform shape).
%In the case of the EEGNet, the filters are simply the 
While this layer typically operates on a single scale, which is determined by the sampling rate, the layer may also operate on multiple scales, such as in the case of the MSA-CNN model.
%To this end, we track the activations of the temporal module relating to the filters in the first temporal convolution. 
%Note that this requires to retain the filter dimension for subsequent temporal convolutions, which would otherwise be averaged over.
The filters of these convolutions capture distinct frequencies, which are determined by their Fourier transform and their scale.
For multi-scale configurations, frequencies at different scales may overlap and thus require unification. 
The amplitude values of the Fourier components are then averaged across filters to yield the final amplitude spectrum.
%using an assignment matrix $\mathbf{S}_\mathrm{unif}$. 
This spectrum is then compared with the between-class spectral variation in the data, which serves as an indicator of classification-relevant spectral information.
An overview of our proposed method for uni-scale (a--b) and multi-scale (c--e) configurations is presented in Figure~\ref{fig:method}.
%, applied to the MSA-CNN~\cite{goerttler2025msacnnlightweightmultiscalecnn},
%In the case of the MSM, the filters are grouped by scale, which also scales the frequencies.
%Finally, the filters' frequency patterns allow to map filter activations to frequency activations using assignment matrices.
%We retrieve the spectral activation pattern evoked by an input EEG channel.
%spectral range that these convolutions cover is determined by their scale. Therefore, the spectral information that each filter can capture depends both on the spectral components of the weights as well as the scale.
%Lastly, the activations of each filter are mapped to activations of each frequency using assignment matrices based on the filter weights, allowing us to retrieve the spectral activation pattern each channel evokes.

\subsection{EEGNet Architecture}
The EEGNet is a lightweight CNN architecture introduced by Lawhern \textit{et al.}~\cite{lawhern2018eegnet}. The model comprises three convolutional layers arranged in two blocks as well as a classification layer. The first block consists of a temporal convolutional layer and a depthwise spatial convolutional layer. 
The second block consists of a separable convolution, which comprises a depthwise temporal convolution and a point-wise convolution.
Each convolutional layer is followed by batch normalisation, while each block ends with a sequence of exponential linear unit, average pooling and dropout. The classification layer consists of a fully connected layer and a softmax operation. 
Note that the spatial layer in the first block and the second block correspond to the higher-order feature extraction block in Figure~\ref{fig:method}(a).
For consistency, the current study uses the same model parameters as the original study. %%% can be deleted if space issue

\subsection{Multi-Scale and Attention Convolutional Neural Network (MSA-CNN)}
The MSA-CNN~\cite{goerttler2025msacnnlightweightmultiscalecnn} is shown in Figure~\ref{fig:method}(c). To begin with, a multi-scale convolution and a scale-integration convolution extract spectral features for each channel.
%The MSM comprises two temporal convolutions, namely 
Subsequently, the channels are combined using a global spatial convolution. The time-resolved features are then passed to an attention-based Temporal Context Module (TCM), which captures long-range temporal dependencies. The TCM combines multi-head attention with a feed-forward neural network.
%(for more details regarding the TCM, see~\cite{goerttler2025msacnnlightweightmultiscalecnn}). 
Finally, the resulting features are averaged over time and classified using a fully connected layer.

\iffalse
\begin{figure}[tbh]
    \centering
    \resizebox{1.035\columnwidth}{!}{%
    \input{graphics/MSM.tikz}%
    }
    \caption{Illustration of the Multi-Scale Module (MSM) with four scales~\cite{goerttler2025msacnnlightweightmultiscalecnn}. The first temporal convolution extracts low-level features on separate scales. A complementary pooling operation allows the feature maps to be merged. The second temporal convolution integrates the scales.}
    \label{fig:MSM}
\end{figure}
\fi

\subsection{Unification Assignment Matrix}
\label{ssec:unif}
\iffalse
We use two assignment matrices to map a filter position encoding of a vector to scale-wise frequencies, and subsequently to unified frequencies. 
%
Firstly, the \textbf{frequency assignment matrix} $\mathbf{S}_{freq}$ is computed from the filter weight matrices $\mathbf{W}^{(\text{I})}$,..., $\mathbf{W}^{(\text{IV})}\in {\mathbb{R}^{N_{f}\times N_{t}}}$ for scales I-IV of the multi-scale convolution, as shown in Figure~\ref{fig:method}(b). Specifically, the discrete Fourier transform (DFT) is computed up to the Nyquist limit along the temporal dimension. The transformation can be described as a matrix multiplication of the positive-frequency DFT matrix $F_{N_{t}}^+$ with the weight matrix. The frequency assignment matrix is then constructed as a block matrix with the absolute transformed weights:
\begin{align}
\mathbf{S}_\text{freq} = 
\begin{bmatrix}
|F_{N_{t}}^+\mathbf{W}^{(\text{I})}|^\top \phantom{000000000000000000000000}\\
\phantom{00000000}|F_{N_{t}}^+\mathbf{W}^{(\text{II})}|^\top \phantom{0000000000000000}\\
\phantom{0000000000000000}|F_{N_{t}}^+\mathbf{W}^{(\text{III})}|^\top \phantom{00000000}\\
\phantom{000000000000000000000000}|F_{N_{t}}^+\mathbf{W}^{(\text{IV})}|^\top \\
\end{bmatrix}.
\end{align}
Note that the transformed weights are transposed such that the columns correspond to frequencies.
\fi

In many deep learning-based neurophysiology models, the first convolutional layer comprises convolutions on more than one scale~\cite{supratak2017deepsleepnet,jia2020graphsleepnet,eldele2021attention,goerttler2025msacnnlightweightmultiscalecnn,zucchi2025prism}.
This requires the use of a unification assignment matrix $\mathbf{S}_\text{unif}$ for our explainability tool, which assigns the scale-based frequencies to unified frequencies on a single scale. 

The scale-based frequencies are collected in the list $\mathbf{f}_\text{sc}=(f_0^{(\text{I})},f_1^{(\text{I})},...,f_0^{(\text{II})},...,f_{\lfloor N_t/2\rfloor}^{(\text{IV})})$, where the superscript denotes the scale and the subscript indexes the frequency within that scale. We determine the list of unique frequencies $\mathbf{f}_\text{uniq}$ by sorting $\mathbf{f}_\text{sc}$ by magnitude and removing duplicate frequencies. The unnormalised unification assignment matrix is then given by:
\begin{align}
    \widehat{\mathbf{S}}_{\text{unif},ij} = \begin{cases} \mbox{$1$,} & \mbox{if } \mathbf{f}_\text{uniq}[i]=\mathbf{f}_\text{sc}[j], \\ \mbox{$0$,} & \mbox{otherwise.} \end{cases}
\end{align}
The matrix is then normalised using the diagonal 
matrix $\mathbf{D}_{ii}=\sum_j \widehat{\mathbf{S}}_{\text{unif},ij}$. This yields
the full frequency unification matrix $\mathbf{S}_\text{unif}=\mathbf{D}^{-1}\widehat{\mathbf{S}}_\text{unif}$, which transforms the scale-based frequencies into their unified representation as
\begin{align}
    \mathbf{f}_\text{unif}=\mathbf{S}_\text{unif}\, \mathbf{f}_\text{sc}.
\end{align}
%yielding the final assignment matrix 
%summed-row 

\subsection{Filter Spectrum Retrieval}
%The filter spectrum retrieval is illustrated in Figure~\ref{fig:method}(c). 
An overview of the uni- and multi-scale filter spectrum retrieval is given in Figure~\ref{fig:method}(b) and \ref{fig:method}(d), respectively. The filter spectrum is retrieved from the filter weights $\mathbf{W}$ of the trained model. In the case of unimodal filters, where filters are shared across all neurophysiological channels, the weight matrix has shape $K\times L \times S$,  with $K$ denoting the number of output channels, $L$ the length of the filter, and $S$ the number of scales.
%For simplicity, only one channel is described. The input signal is processed through the MSM module using the pre-trained filter weights of the MSA-CNN model. Importantly, the second MSM convolution (scale-integration convolution) is modified to retain the filter dimension by convolving the features without averaging across filter dimension, encoded by convolution channels.
%This ensures that the resulting activations can still be related to the filters of the first temporal convolution.
%, such that the resulting activations can still be related to the multi-scale filters. 
%Firstly, the filter weights of the first convolutional layer are extracted from the trained model.
%Note that the layer can be either unimodal, where a single set of filters is trained and shared across all channels, or multimodal, where a separate set of filters is trained for each channel and modality.
%
Subsequently, the discrete Fourier transform (DFT) is applied to the filter weights along the second, temporal dimension. Only the magnitudes of all positive frequencies are kept for further processing. The resulting feature maps are averaged across the $K$ output channels, producing the filter spectrum in terms of frequency and scale. 
%If the convolutional layer comprises more than one scale, 
For multi-scale configurations, the scales are unified by multiplying the filter spectrum by the assignment matrix $\mathbf{S}_\mathrm{unif}$.
%in order to unify the scales.
%The filter spectrum retrieval for multi-scale configurations is illustrated in Figure~\ref{fig:method}(c). 
%, yielding the final filter spectrum.
%The feature map is then firstly averaged over time and filter channels, resulting in an activation vector sorted by filter position. Secondly, the frequency assignment and unification assignment matrices are multiplied with the activation vector, yielding the final activation vector sorted by frequency.

\subsection{Between-Class Spectral Variation}
This section describes a measure that assesses the spectral information relevant for classification for each channel.
%channel-wise spectral information relevant for classification in the data.
%determine to what extent each frequency varies between sleep stages, relative to each channel. 
To this end, we introduce the between-class spectral variation, which quantifies spectral density variation between class means normalised by within-class variation.
%and is normalised by the within-class variation.
Specifically, we compute the spectral density as the square root of Welch's power spectral density for every time series. The between-class variation is then retrieved by firstly averaging the spectral density across all samples for each class, and subsequently computing the standard deviation between the classes. On the other hand, the within-class variation is retrieved by computing the standard deviation for each class before averaging across classes.
Lastly, we divide the between-class variation by the within-class variation to obtain our final measure of relevant spectral information.
%in a given channel which is available for classification.
%representing the class discriminability with respect to frequency and channel.

\section{Experiments}
\label{sec:results}
%\subsection{Alzheimer's disease}
\subsection{Datasets}
We use the publicly available datasets ISRUC-S3 and Sleep-EDF-20 in this study. 
%, all of which are publicly available. 
The \textbf{ISRUC-S3} was recorded by Khalighi et al. from 10 healthy subjects during sleep~\cite{khalighi2016isruc}. The recordings were divided into 30-second epochs and labelled, resulting in a total of 8,589 annotated samples. For our experiment, we only use channels with sufficient spectral structure, which includes the six referenced electroencephalography (EEG) channels, the two electrooculography (EOG) channels, and the electromyogram  (EMG) channel. The input signals are downsampled to 100\,Hz and preprocessed with a fourth-order low-pass Butterworth filter at 40\,Hz cutoff frequency.

The \textbf{Sleep-EDF-20} dataset is sourced from PhysioBank~\cite{goldberger2000physiobank}. The dataset comprises 20 subjects and has overall 42,308 labelled, artefact-free 30-second samples with a sampling rate of 100\,Hz. 
Channels with spectral structure include two referenced EEG channels (Fpz-Cz and Pz-Oz) and the EOG channel. Similar to the ISRUC-S3 dataset, we preprocessed the data with a 40\,Hz low-pass filter.

\subsection{Model Specifications}
We trained multivariate models for generating the weight matrices for the filter spectrum retrieval, as well as univariate models to assess the single channel performance.
Multivariate models were trained on the entire dataset, while univariate models were trained and evaluated using 10-fold subject-wise cross-validation, with performance measured as mean accuracy.

We adopted the parameter settings for EEGNet as described in the original publication~\cite{lawhern2018eegnet}, with minor modifications. 
Specifically, we reduced the filter size from 64 to 50 due to the lower sampling rates in this study,
%to account for the lower sampling rates of our datasets
such that the detectable spectral resolution of 2\,Hz matches that in the original study. 
The same model parameters were used for both univariate and multivariate inputs across both datasets.
To train the models, we followed the procedure outlined in the original publication. 

The model parameters for the MSA-CNN configurations are listed in Table \ref{tab:hyperparameters}. Both temporal convolutional layers are set to unimodal, meaning that filters are shared across channels.
We tailored the model size to each dataset, using a small configuration for ISRUC-S3 and a large configuration for Sleep-EDF-20. In addition, for the univariate configuration, we replaced the spatial convolution in the third layer with a temporal convolution.
All models were trained for 100 epochs using the Adam optimizer~\cite{kingma2014adam}, with a learning rate of 0.001 and a batch size of 64. Regularisation was applied using a dropout rate of 0.1 and weight decay of 0.0001. 
%To generate the filter weights used for the filter spectrum retrieval, multivariate models were trained on the entire dataset, while univariate models were trained and evaluated using 10-fold cross-validation, with performance measured as mean accuracy.
%To evaluate single-channel model performance, we used 10-fold cross-validation and computed the mean accuracy.

\begin{table}[]
    \centering
    \caption{Model parameters for MSA-CNN small and MSA-CNN large, based on multivariate or univariate inputs. Kernel sizes are specified by their spatial and temporal dimensions}
    \resizebox{1\columnwidth}{!}{
    \begin{tabular}{llcccc}
\toprule
layer & hyperparameter & \multicolumn{2}{c}{MSA-CNN (small)}& \multicolumn{2}{c}{MSA-CNN (large)}\\ \cmidrule(lr){3-4}\cmidrule(lr){5-6}
& & multivar. & univar. & multivar. & univar. \\
\midrule
MSM I & \# scales &\multicolumn{2}{c}{4}&\multicolumn{2}{c}{4}\\%\multirow{2}*{temporal I}
 & filter size &\multicolumn{2}{c}{1$\times$15} & \multicolumn{2}{c}{1$\times$15}\\%\multirow{2}*{temporal I}
& \# filters / scale & \multicolumn{2}{c}{8}&\multicolumn{2}{c}{8}\\
MSM II & filter size & \multicolumn{2}{c}{1$\times$5} & \multicolumn{2}{c}{1$\times$5} \\
 & \# filters & \multicolumn{2}{c}{16}&\multicolumn{2}{c}{32}\\ %\cdashlinelr{1-8}
 spatial & filter size &$N_{ch}\times 1$&$1\times 5$ & $N_{ch}\times 1$ & $1\times 5$\\
& \# filters &\multicolumn{2}{c}{32}&\multicolumn{2}{c}{64}\\ %\cdashlinelr{1-8}
TCM & embedding dim.&\multicolumn{2}{c}{16}&\multicolumn{2}{c}{32}\\
 & \# heads&\multicolumn{2}{c}{2}&\multicolumn{2}{c}{4}\\
 & \# layers &\multicolumn{2}{c}{1}&\multicolumn{2}{c}{2}\\
\bottomrule
\end{tabular}}
    \label{tab:hyperparameters}
\end{table}

\iffalse
\begin{figure}[tbh]
    \centering
    \resizebox{1.093\columnwidth}{!}{%
    \input{graphics/TCM.tikz}%
    }
    \caption{Temporal context module using multi-head self-attention. Multi-head attention requires the computation of query (blue), key (green), and value (red) maps. The sequence of attention and feed-forward layer (grey area) is carried out $N_{lay}$ times.}
    \label{fig:TCM}
\end{figure}
\fi
\iffalse
\begin{table}[]
    \centering
    \caption{Scale settings and characteristics in the experiment}
    \resizebox{1\columnwidth}{!}{
    \input{tables/scales}}
    \label{tab:scales}
\end{table}
\fi

\subsection{Filter Spectrum vs. Between-Class Variation}
To test our method, we retrieve the filter spectra for the multivariate models trained on both datasets and compare it to the between-class variation of each channel.
Given that filter settings are shared across channels, only one filter spectrum is retrieved for each dataset.

For the EEGNet, a filter size of 50 at a sampling rate of 100\,Hz corresponds to a detectable frequency range of 0\,Hz to 48\,Hz, with a frequency spacing of 2\,Hz.
In the case of the MSA-CNN, the filters operate on multiple scales.
To unify the scales, a unification assignment matrix was computed using the pooling settings shown in Figure~\ref{fig:method}(a). Based on the filter size setting of 15 and a sampling rate of 100\,Hz, the unified frequencies span the range 0-46.7\,Hz with a maximal frequency spacing of 6.7\,Hz.
%spectral activations in the MSM and the between-class variation for the 9 channels in the ISRUC-S3 dataset and the 3 channels in the Sleep-EDF-20 dataset. Table \ref{tab:scales}

\begin{figure*}[t]
    \centering
    %includegraphics[width=1\columnwidth]{figures/activations.pdf}    
    \begin{tikzpicture}
        %\sffamily
        \node[inner sep=0, outer sep=0] (image) at (0,0) {\includegraphics[width=0.97\textwidth]{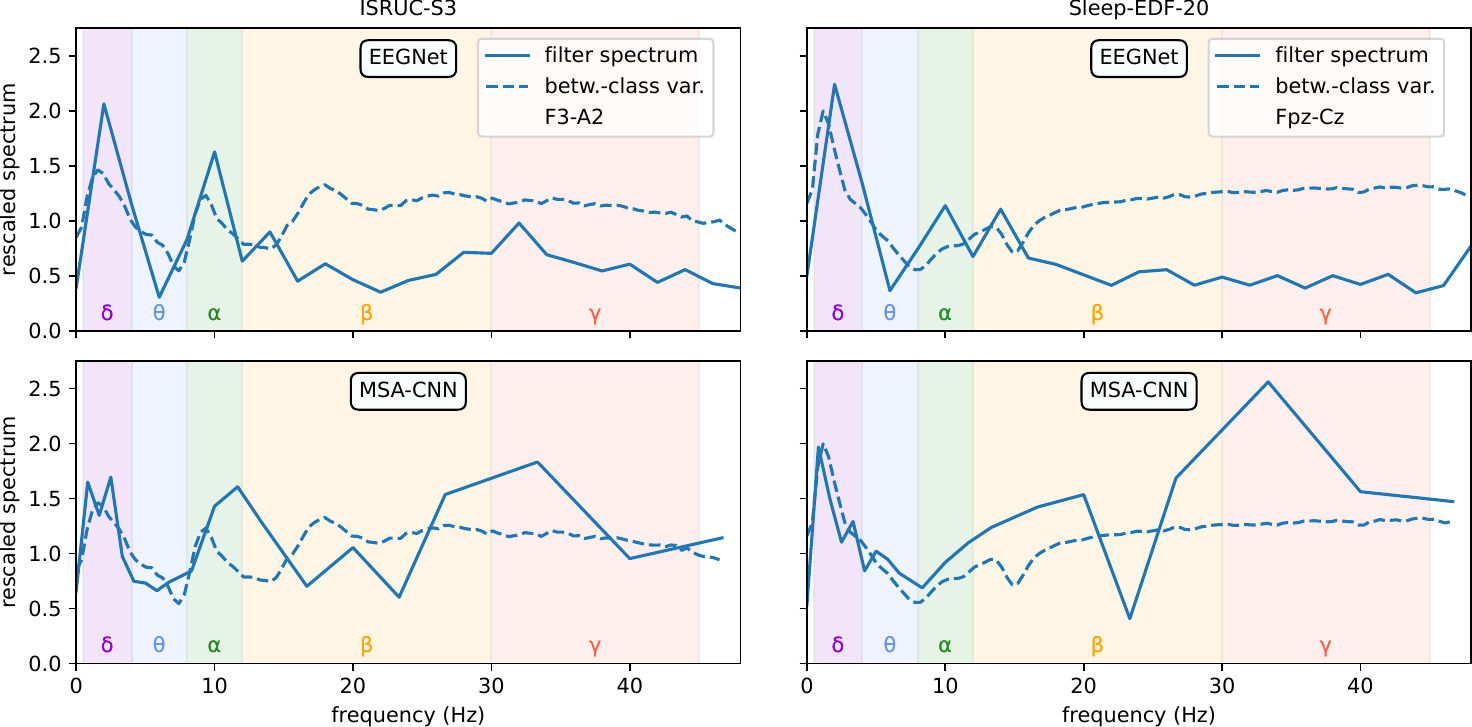}};
        \node[inner sep=0, outer sep=0] (caption) at (-7.1-0.4-0.4,3.75-0.1){\textsf{(a)}}; 
        \node[inner sep=0, outer sep=0] (caption) at (-7.1-0.4-0.4,0.125-0.1){\textsf{(c)}};
        \node[inner sep=0, outer sep=0] (caption) at (0.855-0.4,3.75-0.1){\textsf{(b)}}; 
        \node[inner sep=0, outer sep=0] (caption) at (0.855-0.4,0.125-0.1){\textsf{(d)}};
    \end{tikzpicture}
    \caption{Rescaled filter spectrum (solid lines) for EEGNet (a,b) and MSA-CNN (c,d) on datasets ISRUC-S3 (a,c) and Sleep-EDF-20 (b,d). The spectra are overlaid with the rescaled between-class variation of the frontal channels F3–A2 (ISRUC-S3) and Fpz–Cz (Sleep-EDF-20), shown as dashed lines.
    %Typical frequency bands ($\alpha$, $\beta$, $\gamma$, $\delta$, and $\theta$) are blended in.
    Shaded regions along the frequency axis indicate the standard EEG frequency bands $\delta$ (0.5–4\,Hz), $\theta$ (4–8\,Hz), $\alpha$ (8–12\,Hz), $\beta$ (12–30\,Hz), and $\gamma$ (30–45\,Hz)~\cite{niedermeyer2005electroencephalography}
    At lower frequency bands ($\delta$, $\theta$ and $\alpha$), filter spectra and between-class variations align. At higher frequency bands ($\beta$ and $\gamma$), filter spectra align across datasets.}
    \label{fig:curve}
\end{figure*}

Figure~\ref{fig:curve}(a--b) shows the filter spectrum for the EEGNet model for datasets ISRUC-S3 (a) and Sleep-EDF-20 (b), while Figure~\ref{fig:curve}(c--d) shows the spectrum for the MSA-CNN model for both datasets.
The filter spectrum is overlaid with the between-class variation spectra of the EEG channels F3-A2 (ISRUC-S3) and Fpz-Cz (Sleep-EDF-20), which are commonly used in univariate configurations~\cite{pham2023automatic,eldele2021attention}.
All spectra were rescaled by dividing by the standard deviation in the frequency range 0.5-12\,Hz, which corresponds to the combined lower EEG frequency bands $\delta$ (0.5–4\,Hz), $\theta$ (4–8\,Hz), and $\alpha$ (8–12\,Hz)~\cite{niedermeyer2005electroencephalography}. For both models, the results indicate substantial similarity between filter spectrum and between-class variation in these lower frequency bands on both datasets. For example, both models capture a peak in between-class variation in the $alpha$ band on the ISRUC-S3 dataset, whereas on the Sleep-EDF-20 dataset they capture a plateau.
This result indicates that the model uses mainly lower frequencies for spectral information-based classification.
On the other hand, the filter spectra align across datasets in the higher frequency bands $\beta$ and $\gamma$, which suggests that higher frequencies may be used to construct complex wave patterns shared across EEG datasets.

\subsection{Spectral Information Extraction Relative to Modality}
\begin{figure}[t]
    \centering
    \begin{tikzpicture}
        %\sffamily
        \node[inner sep=0, outer sep=0] (image) at (0,0) {\includegraphics[width=1\columnwidth]{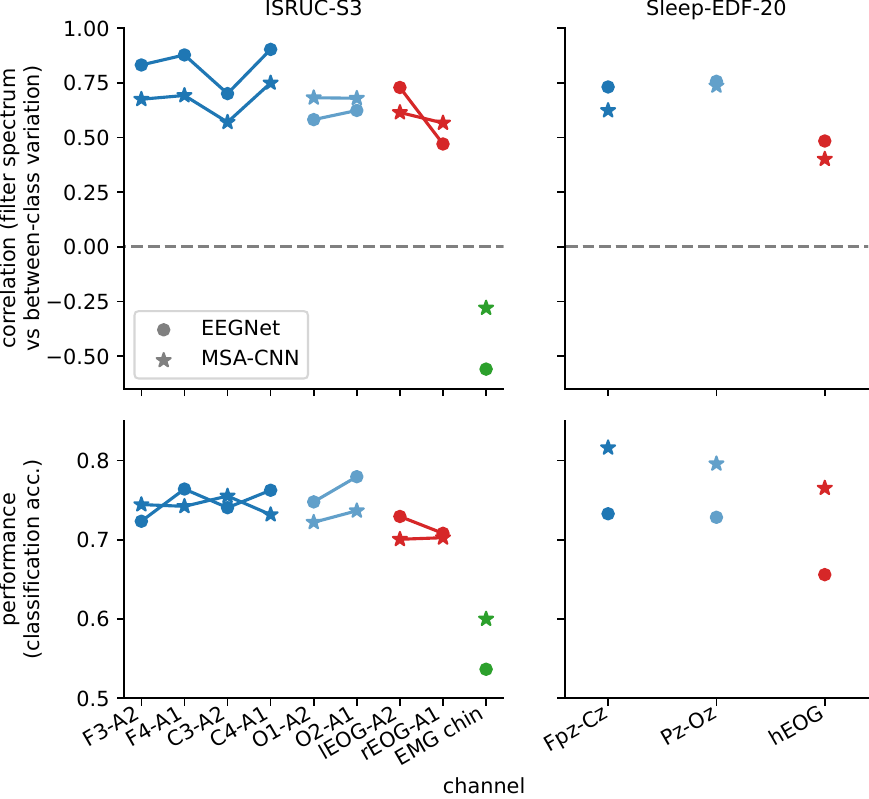}};
        \node (caption) at (-3.65,3.62){\textsf{(a)}}; %+0.75
        \node (caption) at (-3.65,-0.05){\textsf{(c)}};
        \node (caption) at (.92,3.62){\textsf{(b)}}; 
        \node (caption) at (.92,-0.05){\textsf{(d)}};
    \end{tikzpicture}
    \caption{(a,b) Channel-wise Pearson correlation between spectral activation and between-class variation in the lower frequency bands on datasets ISRUC-S3 (a) and Sleep-EDF-20 (b). Generally, EEG channels (blue) exhibit the highest similarity, followed by EOG (red) and EMG (green). (c,d) Channel-wise classification performance for the two respective datasets. 
    Channel performance approximately matches the correlations in a and b.\vspace{-0.2cm}
    }
    \label{fig:correlation}
\end{figure}

To understand the specific role of channel and modality in extracting spectral information, we compute the Pearson correlation of filter spectrum and between-class variation in the lower frequency bands ($\delta$, $\theta$ and $\alpha$) across channels, as shown in Figure~\ref{fig:correlation}(a,b). We find that the correlation is highest for EEG, followed by EOG and EMG, irrespective of the model or dataset.
This pattern is less pronounced for the ISRUC-S3 dataset, where EOG channels correlate almost as strongly with filter spectrum as EEG channels.
There is also a within-modality discrepancy between frontal and occipital EEG channels: While for ISRUC-S3, frontal (blue) and occipital (light blue) channels correlate equally with filter spectrum, for Sleep-EDF-20 occipital channels correlate stronger than frontal channels.

We validate these results by comparing them to the univariate performance for each channel and modality. 
%The similarity is measured as Pearson correlation and as the average dot product, normalised by channel number.
The univariate performances, shown in Figure~\ref{fig:correlation}(c,d), exhibit a similar pattern, with EEG outperforming EOG and EMG.
This matching between correlation and performance indicates that both models are capable of prioritising more informative channels by optimising the retrieval of spectral information for these channels. Given the unimodal configuration of the first layer in both models, this prioritisation comes at the cost of neglecting less informative modalities.

%indicate that, in both datasets, the similarity is highest for frontal EEG channels, followed by occipital EEG channels and then EOG channels. For the EMG channel in the ISRUC-S3 dataset, the two similarity measures yield mixed results.
%Remarkably, channels with higher similarity also exhibit higher single-channel performance. 
%The results highlight the role of spectral information for classification, and show how the model naturally favours the more distinct channels in a multi-channel setting.

%relative to frequency channel-wise across the two datasets on the one hand, and the data-based within-class variations on the other hand.

%Figure xxx shows the 
%compare our retrieved network activations 
%with frequency-based class discriminablity in the data.

\section{Conclusion}
\label{sec:discussion}
This study introduced a tool to explain the role of spectral information processing in CNN-based classification models. We tested the tool on two multivariate sleep stage classification models across two datasets. For both models, we found that the convolutional filter spectrum aligns with the spectral information available in the data at lower frequencies, which indicates that the models optimise spectral information extraction at these frequencies.
Note that the available spectral information depends on the dataset and application. For example, ISRUC-S3 shows greater between-class variation in the $\alpha$ wave range than Sleep-EDF-20, likely reflecting differences in EEG channel configuration and system characteristics.
%TODO
The observed alignment highlights the way in which convolutional layers can function as spectral feature extractors~\cite{rippel2015spectral}.

At higher frequencies, filter spectrum and data-derived spectral information diverge, while filter spectra align across datasets.
%a similarity of filter spectra across datasets and discrepancies between filter spectrum and between-class variation within a dataset, 
This suggests that the filters capture the general morphology of EEG waves rather than spectral information at these frequencies. 
The capability to capture both spectral and morphological features is in contrast to spectral density extractors such as Welch's method, which are limited to spectral features. 
%This highlights the superiority of convolutional layers over classical spectral density extraction when processing raw neurophysiological input signals.
Altogether, the explainability tool provided insight into the patterns of information processing relative to frequency, which were strikingly similar between the two models.
In general, such similarity suggests that the models are functionally comparable.
%In general, these patterns depend on the specific task or application and
%Generally, these patterns depend on the specific task or application. Similar patterns across models indicate that these models are functionally similar.
%The similarity 
%, which were strikingly similar between the two models.
%while divergence indicates that two models are functionally different.
%this tool therefore helps: find out if two models are effectively the same.

In a second step, we used the explainability tool to analyse channel importance.
We found that both models prioritise the more informative channels at the expense of less informative ones. Apart from providing insight into model functioning, this finding may be leveraged for performance optimisation in the following ways:
First, channel selection for single- and multi-channel applications can be performed with only a single trained multivariate model.
Second, comparisons between filter spectrum and between-class variation can be used to balance the frequency scales for optimal extraction of spectral information. 
Third, large differences in between-class variation can signal the need to perform feature extraction separately for each modality.
Together, these examples underscore the value of our method for model improvement specifically, and the value of explainability for healthcare applications more generally.

Nevertheless, several limitations should be noted. First, the correlation estimates may be unreliable due to the limited number of data points used in their computation. In addition, the comparison between correlation estimates and performance for channel importance analysis was only assessed visually. Lastly, the relatively small size of the datasets used in this study may limit the transferability of our model~\cite{supratak2023quantifying}. 
%Validation only visually In addition, the datasets are only small, whcih .... 
%Next, the .... is very limited
%On the other hand, 
In future research, we will improve the channel importance analysis by incorporating channel ablation experiments.
In addition, we plan to test our explainability tool on larger and more diverse datasets.
%incorporate channel ablation experiments to more directly assess the direct contribution of each channel.
%Lastly, it has been shown that ....~\cite{supratak2023quantifying}
%the specifics of datasets not taken into account. Future research will map differecnes ...
%
%asdf
%asdf
%asdfa asdf asdf asdf asdf asdf adsf

%------------------------------------------------------------------------

\section*{Acknowledgements}

Stephan Goerttler was supported by the A*STAR Research Attachment Program (ARAP).
%------------------------------------------------------------------------

%\newpage
%\clearpage
%\balance
% \parskip=4pt
% \setlength{\bibsep}{2.5pt plus 0ex}
\setlength{\bibsep}{2.5pt plus 0ex}
%\baselineskip10pt
%\parskip=2.5pt
%\vspace{-1.cm}
\footnotesize
\bibliographystyle{IEEEbibSPMB}
\bibliography{IEEEabrv,IEEESPMB}

\end{document}